\documentstyle[12pt]{article}

\font\twlgot =eufm10 scaled \magstep1 \font\egtgot =eufm8
\font\sevgot =eufm7 \font\twlmsb =msbm10 scaled \magstep1
\font\egtmsb =msbm8 \font\sevmsb =msbm7

\newfam\gotfam
\def\pgot{\fam\gotfam\twlgot}
\textfont\gotfam\twlgot \scriptfont\gotfam\egtgot
\scriptscriptfont\gotfam\sevgot
\def\got{\protect\pgot}
\newfam\msbfam
\textfont\msbfam\twlmsb \scriptfont\msbfam\egtmsb
\scriptscriptfont\msbfam\sevmsb
\def\Bbb{\protect\pBbb}
\def\pBbb{\relax\ifmmode\expandafter\Bb\else\typeout{You cann't use
Bbb in text mode}\fi}
\def\Bb #1{{\fam\msbfam\relax#1}}

\newcommand{\ccG}{{\got g}}

\def\thebibliography#1{\section*{References}\list
  {[\arabic{enumi}]}{\settowidth\labelwidth{#1}\leftmargin\labelwidth
    \advance\leftmargin\labelsep
    \usecounter{enumi}}
    \def\newblock{\hskip .11em plus .33em minus .07em}
    \sloppy\clubpenalty4000\widowpenalty4000
    \sfcode`\.=1000\relax}

\def\op#1{\mathop{\fam0 #1}\limits}

\newcommand{\id}{{\rm Id\,}}

\newcommand{\beq}{\begin{equation}}
\newcommand{\eeq}{\end{equation}}
\newcommand{\ben}{\begin{eqnarray}}
\newcommand{\een}{\end{eqnarray}}
\newcommand{\be}{\begin{eqnarray*}}
\newcommand{\ee}{\end{eqnarray*}}
\newcommand{\bea}{\begin{eqalph}}
\newcommand{\eea}{\end{eqalph}}

\newcommand{\cH}{{\cal H}}

\newcommand{\bb}{{\bf 1}}

\newcommand{\Is}{{\rm Aut}}

\newcommand{\al}{\alpha}

\newcommand{\bt}{\beta}
\newcommand{\dl}{\delta}
\newcommand{\la}{\lambda}

\newcommand{\f}{\phi}

\newcommand{\m}{\mu}

\newcommand{\th}{\theta}

\newcommand{\lng}{\langle}
\newcommand{\rng}{\rangle}

\newcommand{\si}{\sigma}
\newcommand{\Si}{\Sigma}

\newcommand{\wh}{\widehat}

\newcommand{\dr}{\partial}
\newcommand{\ar}{\op\longrightarrow}
\newcommand{\ot}{\otimes}

\newcounter{theorem}
\newcounter{remark}
\newcounter{proposition}
\newcounter{lemma}
\newcounter{corollary}
\newcounter{definition}

\def\theremark{\arabic{remark}}

\def\thedefinition{\arabic{definition}}

\newenvironment{theo}{\refstepcounter{definition} \medskip
\noindent{\bf Theorem \thedefinition.}}{\medskip}

\newcommand{\mar}[1]{}

\begin{document}
\hbox{}

\begin{center}

{\large\bf MATHEMATICAL MODELS OF SPONTANEOUS SYMMETRY BREAKING}
\bigskip
\bigskip

{\sc G.SARDANASHVILY}

{\it Department of Theoretical Physics, Moscow State University,
117234, Moscow, Russia }

\end{center}
\bigskip
\bigskip

\begin{small}

\noindent {\bf Abstract}. The Higgs mechanism of mass generation
is the main ingredient in the contemporary Standard Model and its
various generalizations. However, there is no comprehensive theory
of spontaneous symmetry breaking. We summarize the relevant
mathematical results characterizing spontaneous symmetry breaking
in algebraic quantum theory, axiomatic quantum field theory, group
theory, and classical gauge theory.

\end{small}

\bigskip
\bigskip

The Higgs mechanism of mass generation is the main ingredient in
the contemporary Standard Model of high energy physics and its
various generalizations. The key point of this mechanism is
interaction of particles and fields with a certain multiplet of
Higgs fields associated to some gauge symmetry group and
possessing nonzero vacuum expectations. The latter is treated as
spontaneous symmetry breaking. The Higgs mechanism has been
extended to different gauge theories including SUSY gauge theory
\cite{chams,nevz,nill,wess}, noncommutative field theory
\cite{dub,mel,ohl}, gravitation theory
\cite{ald,chams2,kakush,kirsch,lecl}, the gauge
Landau--Ginzburg--Higgs theory \cite{arak,govar}. In quantum field
theory, spontaneous symmetry breaking phenomena imply that a
physical vacuum is not the bare Fock one, but it possesses nonzero
physical characteristics and, thus, can interact with particles
and fields.

At present, there is no comprehensive theory of spontaneous
symmetry breaking. In order to attract attention to this problem,
let us summarize the relevant mathematical results characterizing
spontaneous symmetry breaking in algebraic quantum theory,
axiomatic quantum field theory, group theory, and classical gauge
theory.

\section{Algebraic quantum theory}

In algebraic quantum theory, a quantum system is characterized by
a topological involutive algebra $A$ and positive continuous forms
$f$ on $A$. If $A$ is a Banach algebra admitting an approximate
identity (in particular, a $C^*$-algebra), the well-known
Gelfand--Naimark--Segal (GNS) representation theorem associates to
any positive continuous form on an algebra $A$ its cyclic
representation by bounded (continuous) operators in a Hilbert
space \cite{dixm}. There exist different extensions of this GNS
representation theorem \cite{book05}. For instance, axiomatic
quantum field theory (QFT) deals with unnormed unital topological
involutive algebras. Given such an algebra $A$, the GNS
representation theorem associates to a state $f$ of $A$ a strongly
cyclic Hermitian representation $(\pi_f,\th_f)$ of $A$ by an
$Op^*$- algebra $\pi(A)$ of unbounded operators in a Hilbert space
such that $f(a)=\lng \pi(a)\th_f|\th_f\rng$, $a\in A$
\cite{hor,schmu}.

Given a topological involutive algebra $A$ and its representation
$\pi$ in a Hilbert space $E$, one speaks about spontaneous
symmetry breaking if there are automorphisms of $A$ which do not
admit a unitary representation in $E$.

Recall that an automorphism $\rho$ of $A$ possesses a unitary
representation in $E$ if there exists a unitary operator $U_\rho$
in $E$ such that
\mar{081}\beq
\pi(\rho(a))=U_\rho\pi(a)U_\rho^{-1}, \qquad a\in A. \label{081}
\eeq
A problem is that such a representation is never unique. Namely,
let $U$ and $U'$ be arbitrary unitary elements of the commutant
$\pi(A)'$ of $\pi(A)$. Then $UU_\rho U'$ also provides a unitary
representation of $\rho$. For instance, one can always choose
phase multipliers $U=\exp(i\al)\bb\in U(1)$. A consequence of this
ambiguity is the following.

Let $G$ be a group of automorphisms of an algebra $A$ whose
elements $g\in G$ admit unitary representations $U_g$ (\ref{081}).
The set of operators $U_g$, $g\in G$, however need not be a group.
In general, we have
\be
U_g U_{g'}=U(g,g')U_{gg'}U'(g,g'), \qquad U(g,g'),U'(g,g')\in
\pi(A)'.
\ee
If all $U(g,g')$ are phase multipliers, one says that the unitary
operators $U_g$, $g\in G$, form a projective representation
$U(G)$:
\be
 U_g U_{g'}=k(g,g')U_{gg'}, \qquad g,g'\in G,
\ee
of a group $G$ \cite{cass,varad}. In this case, the set
$U(1)\times U(G)$ becomes a group which is a central
$U(1)$-extension
\mar{084}\beq
\bb\ar U(1)\ar U(1)\times U(G)\ar G\ar \bb \label{084}
\eeq
of a group $G$. Accordingly, the projective representation
$\pi(G)$ of $G$ is a splitting of the exact sequence (\ref{084}).
It is characterized by $U(1)$-multipliers $k(g,g')$ which form a
two-cocycle
\mar{085}\beq
k(\bb,g)=k(g,\bb)=\bb, \qquad
k(g_1,g_2g_3)k(g_2,g_3)=k(g_1,g_2)k(g_1g_2,g_3) \label{085}
\eeq
of the cochain complex of $G$ with coefficients in $U(1)$
\cite{book05,mcl}. A different splitting of the exact sequence
(\ref{084}) yields a different projective representation $U'(G)$
of $G$ whose multipliers $k'(g,g')$ form a cocycle equivalent to
the cocycle (\ref{085}). If this cocycle is a coboundary, there
exists a splitting of the extension (\ref{084}) which provides a
unitary representation of a group $G$ of automorphisms of an
algebra $A$ in $E$.

For instance, let $B(E)$ be the $C^*$-algebra of all bounded
operators in a Hilbert space $E$. Any automorphisms of $B(E)$ is
inner and, consequently, possesses a unitary representation in
$E$. Since the commutant of $B(E)$ reduces to scalars, the group
of automorphisms of $B(E)$ admits a projective representation in
$E$, but it need not be unitary.

One can say something if $A$ is a $C^*$-algebra and its GNS
representations are considered \cite{brat,dixm,book05}.

\begin{theo} \mar{t1} \label{t1} Let $f$ be a state of a $C^*$-algebra $A$ and
$(\pi_f,\th_f,E_f)$ the corresponding GNS representation of $A$.
An automorphism $\rho$ of $A$ defines a state
\mar{0810}\beq
(\rho f)(a)=f(\rho(a)), \qquad a\in A, \label{0810}
\eeq
of $A$ such that the carrier space $E_{\rho f}$ of the
corresponding GNS representation $\pi_{\rho f}$ is  isomorphic to
$E_f$.
\end{theo}

It follows that the representations $\pi_{\rho f}$ can be given by
operators $\pi_{\rho f}(a)=\pi_f(\rho(a))$ in the carrier space
$E_f$ of the representation $\pi_f$, but these representations
fail to be equivalent, unless an automorphism $\rho$ possesses a
unitary representation (\ref{081}) in $E_f$.

\begin{theo} \mar{t2} \label{t2}
If a state $f$ of a $C^*$-algebra $A$ is stationary
\mar{082}\beq
f(\rho(a))=f(a), \qquad a\in A, \label{082}
\eeq
with respect to an automorphism $\rho$ of $A$, there exists a
unique unitary representation $U_\rho$ (\ref{0810}) of $\rho$ in
$E_f$ such that
\mar{087}\beq
U_\rho\th_f=\th_f. \label{087}
\eeq
\end{theo}

A topological group $G$ is called a strongly (resp. uniformly)
continuous group of automorphisms of a $C^*$-algebra $A$ if there
is its continuous monomorphism to the group $\Is(A)$ of
automorphisms of $A$ provided with the strong (resp. normed)
operator topology, and if its action on $A$ is separately
continuous. An infinitesimal generator $\dl$ of a strongly
continuous one-parameter group $G(\Bbb R)$ of automorphisms of a
$C^*$-algebra $A$ is an unbounded derivation of $A$
\cite{brat,pow}. This derivation is bounded iff a group $G(\Bbb
R)$ is uniformly continuous.

\begin{theo} \mar{t20} \label{t20} If a one-parameter group $G(\Bbb R)$
of automorphisms of a $C^*$-algebra $A$ is uniformly continuous,
any representation $\pi$ of $A$ in a Hilbert space $E$ yields the
unitary representation of the group $G(\Bbb R)$ in $E$:
\mar{spr579}\beq
\pi(g(t))=\exp(-it\cH), \qquad \pi(\dl(a))=-i[\cH,\pi(a)], \qquad
a\in A, \label{spr579}
\eeq
where $\cH\in \pi(A)''$ is a bounded self-adjoint operator in $E$
\cite{brat,book05}.
\end{theo}

A problem is that a $C^*$-algebra need not admit nonzero bounded
derivations. For instance, no commutative $C^*$-algebra possesses
bounded derivations. Given a strongly continuous one-parameter
group $G(\Bbb R)$ of automorphisms of a $C^*$-algebra $A$, a
representation of $A$ need not imply a unitary representation
(\ref{spr579}) of this group, unless the following sufficient
condition holds.

\begin{theo} \mar{t3} \label{t3}
Let $f$ be a state of a $C^*$-algebra $A$ such that
\be
|f(\dl(a))|\leq\la[f(a^*a) + f(aa^*)]^{1/2}
\ee
for all $a\in A$ and some positive number $\la$, and let $(\pi_f,
\th_f)$ be the corresponding GNS representation of $A$ in a
Hilbert space $E_f$. Then there exist a self-adjoint operator
$\cH$ on a domain $D\subset \pi_f(A)\th_f$ in $E_f$ and a strongly
continuous unitary representation (\ref{spr579}) of $G(\Bbb R)$ in
$E_f$.
\end{theo}

For instance, any strongly continuous one-parameter group of
automorphisms of a $C^*$-algebra $B(E)$ possesses a unitary
representation in $E$.

\begin{theo} \mar{t4} \label{t4}
Let $G$ be a strongly continuous group of automorphisms of a
$C^*$-algebra $A$, and a state $f$ of $A$ be stationary for $G$.
Then there exists a unique unitary representation of $G$ in $E_f$
whose operators obey the equality (\ref{087}).
\end{theo}

\section{Axiomatic QFT}

In axiomatic QFT, the spontaneous symmetry breaking phenomenon is
described by the Goldstone theorem \cite{bogol}.

There are two main algebraic formulation of QFT. In the framework
of the first one, called local QFT, one associates to a certain
class of subsets of a Minkowski space a net of von Neumann, $C^*$-
or $Op^*$-algebras which obey certain axioms
\cite{araki,haag,hor}. Its inductive limit is called either a
global algebra (in the case of von Neumann algebras) or a
quasilocal algebra (for a net of $C^*$-algebras).

In a different formulation of algebraic QFT, quantum field
algebras are tensor algebras. Let $Q$ be a nuclear space. Let us
consider the direct limit
\mar{x2}\beq
A_Q=\wh\ot Q =\Bbb C \oplus Q \oplus Q\wh\ot Q\oplus \cdots
Q^{\wh\ot n}\oplus\cdots \label{x2}
\eeq
of the vector spaces
\be
\wh\ot^{\leq n} Q =\Bbb C \oplus Q \oplus Q\wh\ot Q\cdots \oplus
Q^{\wh\ot n},
\ee
where $\wh\ot$ is the topological tensor product with respect to
Grothendieck's topology. One can show that, provided with the
inductive limit topology, the tensor algebra $A_Q$ (\ref{x2}) is a
unital nuclear barreled b-algebra \cite{belang,igu}. Therefore,
one can apply to it the GNS representation theorem. Namely, if $f$
is a positive continuous form on $A$, there exists a unique cyclic
representation $\pi_f$ of $A$ in a Hilbert space by operators on a
common invariant domain $D$ \cite{igu}. This domain can be
topologized to conform a rigged Hilbert space such that all the
operators representing $A$ are continuous on $D$. Herewith, a
linear form $f$ on $A_Q$ is continuous iff the restriction of $f$
to each $\wh\ot^{\leq n} Q$ is so \cite{trev}.

In algebraic QFT, one usually choose $Q$ the Schwartz space of
functions of rapid decrease. For the sake of simplicity, we here
restrict our consideration to real scalar fields. One associates
to them the Borchers algebra
\mar{qm801}\beq
A=\Bbb R\oplus RS^4\oplus RS^8\oplus\cdots, \label{qm801}
\eeq
where $RS^{4k}$ is the nuclear space of smooth real functions of
rapid decrease on $\Bbb R^{4k}$ \cite{borch,hor}. It is the real
subspace of the space $S(\Bbb R^{4k})$ of smooth complex functions
of rapid decrease on $\Bbb R^{4k}$. Its topological dual is the
space $S'(\Bbb R^{4k})$ of tempered distributions (generalized
functions). Since the subset $\op\ot^kS(\Bbb R^4)$ is dense in
$S(\Bbb R^{4k})$, we henceforth identify $A$ with the tensor
algebra $A_{RS^4}$ (\ref{x2}). Then any continuous positive form
on the Borchers algebra $A$ (\ref{qm801}) is represented by a
collection of tempered distributions $\{W_k\in S'(\Bbb R^{4k})\}$
such that
\be
f(\psi_k)= \int
W_k(x_1,\ldots,x_k)\psi_k(x_1,\ldots,x_k)d^4x_1\cdots d^4x_k,
\qquad \psi_k\in RS^{4k}.
\ee
For instance, the states of scalar quantum fields in the Minkowski
space $\Bbb R^4$ are described by the Wightman functions
$W_n\subset S'(\Bbb R^{4k})$ in the Minkowski space which obey the
Garding--Wightman axioms of axiomatic QFT \cite{bogol,wigh,zin}.
Let us mention the Poincar\'e covariance axiom, the condition of
the existence and uniqueness of a vacuum $\th_0$, and the spectrum
condition. They imply that: (i) the carrier Hilbert space $E_W$ of
Wightman quantum fields admits a unitary representation of the
Poinar\'e group, (ii) the space $E_W$ contains a unique (up to
scalar multiplications) vector $\psi_0$, called the vacuum vector,
invariant under Poincar\'e transformations, (iii) the spectrum of
the energy-momentum operator lies in the closed positive light
cone. Let $G$ be a connected Lie group of internal symmetries
(automorphisms of the Borchers algebra $A$ over $\id \Bbb R^4$)
whose infinitesimal generators are given by conserved currents
$j^k_\m$. One can show the following \cite{bogol}.

\begin{theo} \mar{t5} \label{t5}
A group $G$ of internal symmetries possesses a unitary
representation in $E_W$ iff the Wightman functions are
$G$-invariant.
\end{theo}

\begin{theo} \mar{t6} \label{t6}
A group $G$ of internal symmetries admits a unitary representation
if a strong spectrum condition holds, i.e., there exists a mass
gap.
\end{theo}

As a consequence, we come to the above mentioned Goldstone
theorem.

\begin{theo} \mar{t7} \label{t7}
If there is a group $G$ of internal symmetries which are
spontaneously broken, there exist elements $\f\in E_W$ of zero
spin and mass such that $\lng \f| j^k_\m\psi_0\rng\neq 0$ for some
generators of $G$.
\end{theo}

These elements of unit norm are called Goldstone states. It is
easily observed that, if a group $G$ of spontaneously broken
symmetries contains a subgroup of exact symmetries $H$, the
Goldstone states carrier out a homogeneous representation of $G$
isomorphic to the quotient $G/H$. This fact attracted great
attention to such kind representations.

\section{Nonlinear realizations of Lie algebras and superalgebras}

In a general setting, given a Lie group $G$ and its Lie subgroup
$H$, one can construct an induced representation of a group $G$ on
a space of functions $f$ on $G$ taking values in a carrier space
$V$ of some representation of $H$ such that $f(gh)=h^{-1}f(g)$ for
all $h\in H$, $g\in G$ \cite{col68,mack}. If $G\to G/H$ is a
trivial fiber bundle, there exists its global section whose values
are representatives of elements of $G/H$. Given such a section
$s$, the product $G/H\times V$ can be provided with the particular
induced representation
\mar{g4}\beq
G\ni g:(\si,v) \mapsto (g\si, g_\si v), \qquad
g_\si=s(g\si)^{-1}gs(\si)\in H, \label{g4}
\eeq
of $G$. If $H$ is a Cartan subgroup of $G$, the well known
nonlinear realization of $G$ in a neighborhood of its unit
\cite{col,jos} exemplifies the induced representation (\ref{g4}).
In fact, it is a representation of the Lie algebra of $G$ around
its origin as follows.

The Lie algebra $\ccG$ of a Lie group $G$ containing a Cartan
subgroup $H$ is split into the sum $\ccG={\got f} +{\got h}$ of
the Lie algebra $\got h$ of $H$ and its supplement $\got f$
obeying the commutation relations
\be
[{\got f},{\got f}]\subset \got h, \qquad [{\got f},{\got
h}]\subset \got f.
\ee
In this case, there exists an open neighbourhood $U$ of the unit
of $G$ such that any element $g\in U$ is uniquely brought into the
form
\be
g=\exp(F)\exp(I), \qquad F\in{\got f}, \qquad I\in\got h.
\ee
Let $U_G$ be an open neighbourhood of the unit of $G$ such that
$U_G^2\subset U$, and let $U_0$ be an open neighbourhood of the
$H$-invariant center $\si_0$ of the quotient $G/H$ which consists
of elements
\be
\si=g\si_0=\exp(F)\si_0, \qquad g\in U_G.
\ee
Then there is a local section $s(g\si_0)=\exp(F)$ of $G\to G/H$
over $U_0$. With this local section, one can define the induced
representation (\ref{g4}) of elements $g\in U_G\subset G$ on
$U_0\times V$ given by the expressions
\mar{g5,6}\ben
&& g\exp(F)=\exp(F')\exp(I'), \label{g5}\\
&& g:(\exp(F)\si_0,v)\mapsto (\exp(F')\si_0,\exp(I')v). \label{g6}
\een
The corresponding representation of the Lie algebra $\ccG$ of $G$
takes the following form. Let $\{F_\al\}$, $\{I_a\}$ be the bases
for $\got f$ and $\got h$, respectively. Their elements obey the
commutation relations
\be
[I_a,I_b]= c^d_{ab}I_d, \qquad [F_\al,F_\bt]= c^d_{\al\bt}I_d,
\qquad [F_\al,I_b]= c^\bt_{\al b}F_\bt.
\ee
Then the relation (\ref{g5}) leads to the formulas
\mar{g7,7',8}\ben
&& F_\al: F\mapsto F'=F_\al
+\op\sum_{k=1}l_{2k}[\op\ldots_{2k}[F_\al,F],F],\ldots,F]-
l_n\op\sum_{n=1}
 [\op\ldots_n[F,I'],I'],\ldots,I'], \label{g7}\\
&& \qquad I'=
\op\sum_{k=1}l_{2k-1}[\op\ldots_{2k-1}[F_\al,F],F],\ldots,F],
\label{g7'}\\
&& I_a: F\mapsto F'=
2\op\sum_{k=1}l_{2k-1}[\op\ldots_{2k-1}[I_a,F],F],\ldots,F],
\qquad I'=I_a, \label{g8}
\een
where coefficients $l_n$, $n=1,\ldots$, are obtained from the
recursion relation
\mar{g71}\beq
\frac{n}{(n+1)!}=\op\sum_{i=1}^n\frac{l_i}{(n+1-i)!}. \label{g71}
\eeq
Let $U_F$ be an open subset of the origin of the vector space
$\got f$ such that the series (\ref{g7}) -- (\ref{g8}) converge
for all $F\in U_F$, $F_\al\in\got f$ and $I_a\in\got h$. Then the
above mentioned nonlinear realization of the Lie algebra $\ccG$ in
$U_F\times V$ reads
\mar{g9'}\beq
F_\al: (F,v)\mapsto (F',I'v), \qquad I_a:(F,v)\mapsto (F',I'v),
\label{g9'}
\eeq
where $F'$ and $I'$ are given by the expressions (\ref{g7}) --
(\ref{g8}). In physical models, the coefficients $\si^\al$ of
$F=\si^\al F_\al$ are treated as Goldstone fields.

Nonlinear realizations of many groups especially in application to
gravitation theory have been studied
\cite{ish,kirsch,lecl,tiembl}. Furthermore, SUSY gauge theory
including supergravity  is mainly developed as a Yang--Mills type
theory with spontaneous breaking of supersymmetries
\cite{binet,niew,nill,wess}. For instance, let us mention various
superextensions of the pseudo-orthogonal and Poincar\'e Lie
algebras \cite{aleks2,aur2}. The nonlinear realization of a number
of Lie superalgebras have been obtained \cite{clark,ivanov} in
accordance with the following scheme.

Let $G$ be a Lie supergroup in the category of $G$-supermanifolds
\cite{bart}, and let $\wh H$ be its Lie supersubgroup such that
the even part $\wh{\got h}_0$ of its Lie superalgebra is a Cartan
subalgebra of the Lie algebra $\wh\ccG_0$. With
$F,F',F''\in\wh{\got f}_0$ and $I',I''\in\wh{\got h}_0$, we can
repeat the relations (\ref{g5}), (\ref{g7}) -- (\ref{g8}) as
follows:
\mar{g72,3,4}\ben
&& F''\exp(F)=\exp(F')\exp(I'), \nonumber\\
&& \qquad F'=F''
+\op\sum_{k=1}l_{2k}[\op\ldots_{2k}[F'',F],F],\ldots,F]-
l_n\op\sum_{n=1}
 [\op\ldots_n[F,I'],I'],\ldots,I'], \label{g72}\\
&& \qquad I'=
\op\sum_{k=1}l_{2k-1}[\op\ldots_{2k-1}[F'',F],F],\ldots,F],
\label{g73}\\
&& I''\exp(F)=\exp(F')\exp(I'), \nonumber\\
&& \qquad F'=
2\op\sum_{k=1}fl_{2k-1}[\op\ldots_{2k-1}[I'',F],F],\ldots,F],
\qquad I'=I'', \label{g74}
\een
where coefficients $l_n$, $n=1,\ldots$, are obtained from the
formula (\ref{g71}). Let a superspace $\wh V$ carries out a linear
representation of the Lie superalgebra $\wh{\got h}$. Let $\wh
U_F$ be an open subset of the supervector space $\wh{\got f}_0$
such that the series (\ref{g72}) -- (\ref{g74}) converge for all
$F\in \wh U_F$, $F''\in\wh{\got f}_0$ and $I''\in\wh{\got h}_0$.
Then we obtain the following nonlinear realization of the even Lie
algebra $\wh\ccG_0$ in $\wh U_F\times \wh V$:
\be
F'': (F,v)\mapsto (F',I'v), \qquad I'':(F,v)\mapsto (F',I'v),
\ee
where $F'$ and $I'$ are given by the expressions (\ref{g72}) --
(\ref{g74}).

\section{Classical gauge theory}

If $G$ is a real Lie group and $H$ is its closed (and,
consequently, Lie) subgroup, classical fields taking values in the
quotient space $G/H$ characterize spontaneous breaking phenomena
in classical gauge theory. They are called Higgs fields.

In gauge theory on a principal bundle $P\to X$ with a structure
Lie roup $G$, gauge potentials are identified to principal
connections on $P\to X$. Being equivariant under the canonical
action of $G$ on $P$, these connections are represented by
sections of the quotient $C=J^1P/G$, of the first order jet
manifold $J^1P$ of the principal bundle $P\to X$. It is an affine
bundle coordinated by $(x^\la, a^r_\la)$ such that, given a
section $A$ of $C\to X$, its components $A^r_\la=a^r_\la\circ A$
are coefficients of the familiar local connection form \cite{kob},
i.e., gauge potentials.

In gauge theory on a principal bundle $P\to X$, matter fields are
represented by sections of an associated bundle
\mar{0812}\beq
Y=(P\times V)/G, \label{0812}
\eeq
where $V$ is a vector space which the structure group $G$ acts on,
and the quotient (\ref{0812}) is defined by identification of the
elements $(p,v)$ and $(pg,g^{-1}v)$ for all $g\in G$. Any
principal connection $A$ on $P$  yields an associated linear
connection on the associated bundle (\ref{0812}). Given bundle
coordinates $(x^\m,y^i)$ on $Y$, this connection takes the form
\be
A=dx^\m\ot(\dr_\m + A^r_\m I_r^i\dr_i),
\ee
where $I_r$ are generators of a representation of a group $G$ in
$V$.

Spontaneous symmetry breaking in classical gauge theory occurs if
matter fields carrier out a representation only of some subgroup
$H$ of a group $G$. To describe these fields, one should assume
that the structure group $G$ of a principal bundle $P\to X$  is
reduced to its closed Lie subgroup $H$, i.e., $P$ contains an
$H$-principal subbundle called a $G$-structure
\cite{gor,keyl,nik,sard92,ijgmmp06}. We have the composite bundle
\be
P\ar^{\pi_{P\Si}} P/H\ar X,
\ee
where
\mar{b3194}\beq
 P_\Si=P\ar^{\pi_{P\Si}} P/H \label{b3194}
\eeq
is a principal bundle with the structure group $H$ and
\be
\Si=P/H\ar^{\pi_{\Si X}} X
\ee
is a $P$-associated fiber bundle with the typical fiber $G/H$
which the structure group $G$ acts on on the left.

\begin{theo} \mar{t8} \label{t8}
There is one-to-one correspondence between the reduced
$H$-principal subbundles $P^h$ of $P$ and the global sections $h$
of the quotient bundle $P/H\to X$ \cite{ijgmmp06,ste}.
\end{theo}

These sections are the above mentioned classical Higgs fields.
Given such a section $h$, the corresponding reduced subbundle is
the pull-back
\mar{0820}\beq
P^h=h^*P_\Si=\pi_{P\Si}^{-1}(h(X)) \label{0820}
\eeq
of the $H$-principal bundle (\ref{b3194}) onto $h(X)\subset \Si$.

In general, there is a  topological obstruction to the reduction
of a structure group of a principal bundle to its subgroup. One
usually refers to the following fact.

\begin{theo} \mar{t9} \label{t9}
Any fiber bundle whose typical fiber is diffeomorphic to an
Euclidean space has a global section \cite{ste}. In particular,
any structure group $G$ of a principal bundle is always reducible
to its maximal compact subgroup $H$ since the quotient space $G/H$
is homeomorphic to an Euclidean space.
\end{theo}

For instance, this is the case of the groups $G=GL(n,\Bbb C)$,
$H=U(n)$ and $G=GL(n,\Bbb R)$, $H=O(n)$. In the last case, the
associated Higgs field is a Riemannian metric on $X$.

It should be emphasized that different $H$-principal subbundles
$P^h$ and $P^{h'}$ of a $G$-principal bundle $P$ need not be
isomorphic to each other in general. They are isomorphic over $X$
iff there is a vertical automorphism of a principal bundle $P\to
X$ which sends $P^h$ onto $P^{h'}$ \cite{book00,ijgmmp06}. If the
quotient $G/H$ is diffeomorphic to an Euclidean space (e.g., $H$
is a maximal compact subroup of $G$), all $H$-principal subbundles
of a $G$-principal bundle $P$ are isomorphic to each other
\cite{ste}.

If a structure group $G$ of a principal bundle $P\to X$ is
reducible to its subgroup $H$, one can describe matter fields with
an exact symmetry group $H$ as follows.  Let $Y\to \Si$ be a
vector bundle associated to the $H$-principal bundle $P_\Si$
(\ref{b3194}). It is a composite fiber bundle
\be
Y\ar^{\pi_{Y\Si}} \Si\ar^{\pi_{\Si X}} X.
\ee
Let $h$ be a global section of the fiber bundle $\Si\to X$, i.e.,
a Higgs field. Then the restriction
\mar{S10}\beq
Y_h=h^*Y \label{S10}
\eeq
of the fiber bundle $Y\to\Si$ to $h(X)\subset \Si$ is a subbundle
of the fiber bundle $Y\to X$ which is associated to the reduced
$H$-principal bundle $P^h$ (\ref{0820}). One can think of sections
$s_h$ of the fiber bundle $Y_h\to X$ (\ref{S10}) as being matter
fields in the presence of a Higgs field $h$. Given a different
Higgs field $h'$, matter fields in its presence are described by
sections of a different fiber bundle $Y_{h'}$, which is isomorphic
to $Y_h$ iff the $H$-principal bundles $P^h$ and $P^{h'}$ are
isomorphic. The totality of all the pairs $(s_h,h)$ of matter and
Higgs fields is represented by sections of the fiber bundle $Y\to
X$ as follows \cite{book00}.

\begin{theo} \label{t21} \mar{t21}
Since $Y^h\to X$ is a subbundle of $Y\to X$, any section $s_h$ of
$Y^h\to X$ is a section of $Y\to X$ projected onto a section
$h=\pi_{Y\Si}\circ s_h$ of the fiber bundle $\Si\to X$.
Conversely, a section $s$ of $Y\to X$ is a composition
$s=s_\Si\circ h$ of a section $h=\pi_{Y\Si}\circ s$ of $\Si\to X$
and some section $s_\Si$ of the fiber bundle $Y\to\Si$ whose
restriction to the submanifold $h(X)\subset \Si$ is a section
$s_h$ of $Y_h$.
\end{theo}

Turn now to the properties of gauge fields compatible with
spontaneous symmetry breaking. Given a Higgs field $h$, the fiber
bundle $Y^h\to X$ of matter fields is provided with a connection
associated to a principal connection on the $H$-principal bundle
$P^h$.

\begin{theo} \mar{t10} \label{t10}
Any principal connection $A^h$ on a reduced subbundle $P^h$ of $P$
gives rise to a principal connection on $P$, and yields an
associated connection on $P/H\to X$ such that the covariant
differential $D_{A^h}h$ of $h$ vanishes. Conversely, a principal
connection $A$ on $P$ is projected onto $P^h$ iff $D_Ah=0$
\cite{kob}.
\end{theo}

\begin{theo} \mar{t11} \label{t11}
If the Lie algebra ${\ccG}$ of $G$ is the direct sum ${\ccG} =
{\got h} \oplus {\got m}$ of the Lie algebra ${\got h}$ of $H$ and
a subspace ${\got m}\subset {\ccG}$ such that $ad(g)({\got
m})\subset {\got m}$, $g\in H$, then the pull-back of the ${\got
H}$-valued component of any principal connection on $P$ onto a
reduced subbundle $P^h$ is a principal connection on $P^h$. This
is the case of so-called reductive $G$-structure
\cite{godina03,book00}.
\end{theo}

Gravitation theory exemplifies classical gauge theory with
spontaneously broken symmetries where Dirac spinor fields are
matter fields possessing exact Lorentz symmetries and a
pseudo-Riemannian metric plays the role of a Higgs field
\cite{iva,sard98a,sard02,pref06}.

\end{document}